\documentclass[twocolumn,showpacs,amsmath,amssymb,aps,prl,superscriptaddress,floatfix]{revtex4}
\usepackage{graphicx}
\def\bea{\begin{eqnarray}}
\def\eea{\end{eqnarray}}
\def\be{\begin{equation}}
\def\ee{\end{equation}}  
\def\S{\mbox{\bf S}}
\def\et{{\it et al.}}
\begin{document}
\author{A.\ L\"auchli}
\email{laeuchli@comp-phys.org}
\affiliation{Laboratoire de Physique Th\'eorique, CNRS UMR-5152, Univ.~Paul Sabatier, F-31062 Toulouse, France}
\author{G.\ Schmid}
\affiliation{Theoretische Physik, ETH H\"onggerberg, CH-8093 Z\"urich, Switzerland}
\author{S.\ Trebst}
\affiliation{Theoretische Physik, ETH H\"onggerberg, CH-8093 Z\"urich, Switzerland}
\affiliation{Computational Laboratory, ETH Zentrum, CH-8092 Z\"urich, Switzerland}
\date{\today}
\title{Spin nematics in the bilinear-biquadratic S=1 spin chain}
\begin{abstract}
  We report the existence of an extended critical, nondimerized region in the phase diagram 
  of the bilinear-biquadratic spin-one chain. The dominant power law correlations are 
  ferroquadrupolar {\it i.e.}~spin nematic in character. Another known critical region 
  is also characterized by  dominant quadrupolar correlations, although with a different wave vector. 
  Our results show that  spin nematic correlations play an important role in quantum magnets with 
  spin $S \ge 1$ in regions between antiferromagnetic and ferromagnetic phases.
\end{abstract}
\pacs{75.10.Jm, 75.10.Pq, 75.40.Mg, 75.40.Cx}
\maketitle

\begin{figure}[b]
  \centerline{\includegraphics[width=0.7\linewidth]{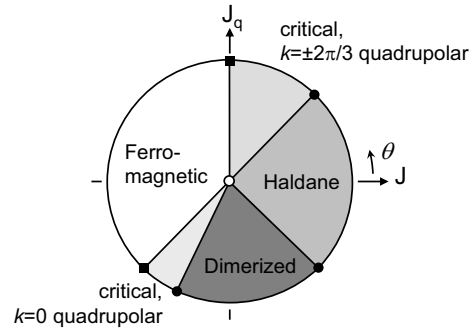}}
  \caption{Phase diagram of the bilinear-biquadratic spin-one chain.
    The new elements are (a) an extended, critical phase with dominant 
    ferroquadrupolar correlations for $-3 \pi/4 < \theta \lesssim -0.67 \pi$ 
    and (b) the characterization of the other critical region $\pi/4 \le \theta < \pi/2$ 
    as having dominant $k=\pm2\pi/3$ quadrupolar correlations.
    \label{fig:PhaseDiagram}}
\end{figure}

The recent experimental demonstration \cite{OrzelGreiner} of the transition to a Mott 
insulating state of atoms in an optical lattice from a superfluid state has opened the
way to novel realizations of effective quantum lattice models with widely tunable control
parameters. Quantum magnetic systems can be realized by spinor atoms in an optical lattice, e.g. 
$^{23}$Na with a total $S=1$ moment. Confining $S=1$ atoms to an optical lattice there
are two scattering channels for identical atoms with total spin $S=0,2$ which can be 
mapped to an effective bilinear and biquadratic spin interaction \cite{Yip,Demler}:
\be
H=\sum_{\langle i,j \rangle} \left[ J \left(\S_{i}\cdot\S_{j} \right) 
  + J_q \ \left(\S_{i} \cdot\S_{j} \right)^2\right],
\label{eqn:biquadHamiltonian}
\ee
where we adopt the standard parametrization $J=\cos \theta$ and $J_q=\sin\theta$. 
In one dimension, the bilinear-biquadratic spin-one model has a rich phase diagram (see Fig.~\ref{fig:PhaseDiagram}), 
which includes the well-known Haldane gap phase \cite{HaldaneConjecture}, a dimerized phase \cite{DimerizedPhase}, 
a ferromagnetic phase and - as we will show in this Letter - two critical spin nematic phases. 
In a pioneering paper \cite{ChubukovPRB} Chubukov suggested the existence  of a gapped, nondimerized phase with 
dominant spin nematic correlations close to the ferromagnetic region of the phase diagram. Subsequent numerical 
work \cite{FathNematic} could however not substantiate this claim and it was therefore believed that the dimerized 
phase would extend up to the ferromagnetic phase boundary. 
Recent quantum Monte Carlo calculations \cite{KawashimaPTP} and field theoretical work \cite{IvanovKolezhuk} 
suggest that this picture might need to be reconsidered, especially in the light of possible experimental
verifications in Bose-Einstein condensate systems \cite{Yip}.

In dimensions higher than one, the bilinear-biquadratic spin-one model is well understood in the relevant
parameter regime. It exhibits a spin nematic phase with a finite spin quadrupole moment in the
groundstate \cite{PapaUnusualPhases,HaradaKawashimaPRB}. 

%%%%%%%%%%%%%%%%%%%%%%%%%%%%%%%%%%%%%%%%%
%% Figure
%%%%%%%%%%%%%%%%%%%%%%%%%%%%%%%%%%%%%%%%%

In this Letter we propose a novel scenario for the phase diagram of the one-dimensional model in between the 
dimerized and the ferromagnetic phases. In particular, we substantiate the existence of an extended critical,
nondimerized region with dominant spin nematic correlations using extensive numerical simulations. We further
characterize a second extended critical phase between the Haldane and the ferromagnetic phase, as also having
prevalent quadrupolar correlations, albeit with a different wavevector. 

The outline of this paper is as follows. In the first part we present various numerical results providing
evidence for an extended critical phase without spontaneous dimerization for $\theta \in (-3\pi/4,-0.67\pi]$.
We show that the quintuplet gap vanishes over this whole interval and that the transition between the dimerized 
and the critical phase is of a generalized Beresinski-Kosterlitz-Thouless (BKT) type. The absence of finite 
dimerization is discussed. We further determine conformal field theory parameters, such as scaling dimensions 
and the central charge for the critical region. Finally we address the second extended critical phase
$\theta \in [+\pi/4,+\pi/2)$, which we exhibit to have prevailing quadrupolar correlations at
wavevectors $\pm 2\pi/3$. 

%% Numerical techniques

{\em Numerical techniques --- } 
Our results constitute the synthesis of different computational methods:
i) Exact diagonalization (ED) of periodic chains up to 18 sites, using both
static and dynamic observables, 
ii) Large scale Density Matrix Renormalization
Group (DMRG) \cite{SRWDmrg} calculations using the finite size algorithm on systems
up to 512 sites while keeping 1000 states,
iii) Strong coupling series expansions \cite{SinghGelfand} around the dimerized
limit up to 10th order.

%% DMRG

\begin{figure}
  \centerline{\includegraphics[width=0.8\linewidth]{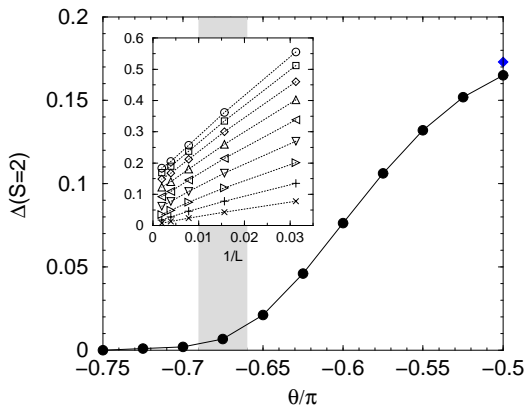}}
  \vspace{-4mm}
  \caption{
    Extrapolated DMRG $S=2$ gaps as a function of $\theta$.
    The gray region denotes the estimated onset of the critical behavior. The
    filled diamond is the exactly known gap result at $\theta=-\pi/2$.
    Inset: finite size extrapolation of the $S=2$ gaps (32 to 512 sites).
    $\theta/\pi$ varies from $-0.5$ to $-0.7$ with decrements of $0.05$ 
    from top to bottom.
    \label{fig:dmrggaptheta}
    \vspace{-5mm} }
\end{figure}

\begin{figure}
  \centerline{\includegraphics[width=0.95\linewidth]{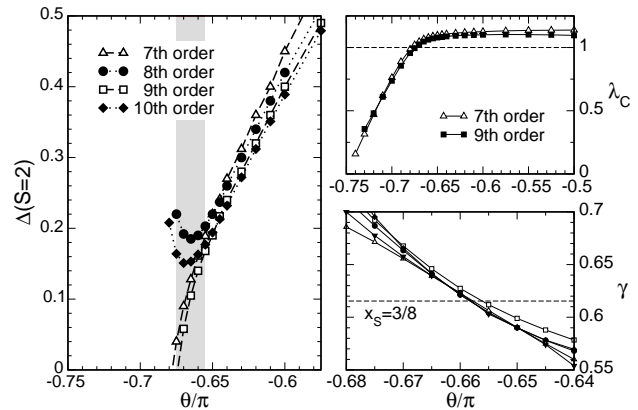}}
  \vspace{-2mm}
  \caption{
    Strong coupling dimer expansion of the $S=2$ gap. 
    Left panel: $S=2$ gap as a function of $\theta$. The gap closes at $\theta_c\approx -0.67 \pi$.
    Upper right panel: Critical value $\lambda_c$ where the gap vanishes. The uniformly coupled 
    chain of interest here corresponds to $\lambda=1$ (dashed line).
    Lower right panel: Critical exponent of the $S=2$ gap opening as a function of the dimerization 
    $(\Delta \sim |\lambda -\lambda_c|^{\gamma})$ calculated by various DLog Pad\'e approximants.
    \label{fig:serieslambda}
    }
\end{figure}

\begin{figure}
  \centerline{\includegraphics[width=0.8\linewidth]{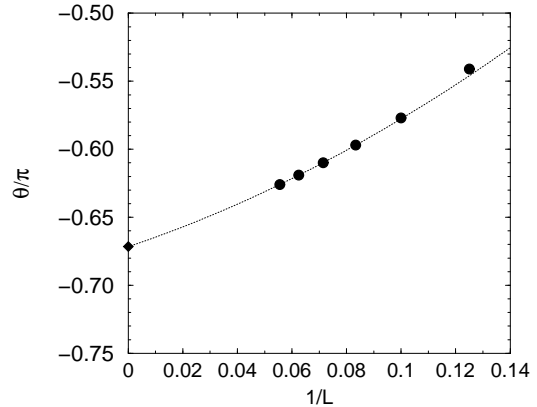}}
  \vspace{-4mm}
  \caption{
    Finite size scaling of the crossing points between the
    singlet gap at momentum $\pi$ and the quintuplet ($S=2$)
    gap at momentum 0. The extrapolation yields a critical 
    value $\theta_c \approx -0.67\pi$ in the thermodynamic limit.
    \label{fig:criticalgaptheta} 
    \vspace{-5mm}}
\end{figure}

{\em Closing of the Gap --- }
We first track the evolution of the energy gap to the first excitation using DMRG
calculations on long open chains. A phase transition is signaled by the closing of the gap.
The lowest excited state on open chains carries spin two for $\theta \in (-3\pi/4,-\pi/2]$ 
\cite{FathNematic}. Our results shown in Fig.~\ref{fig:dmrggaptheta} 
are clearly consistent with a finite gap of the $S=2$ excitation for $\theta \ge -0.65 \pi$. 
However, for $\theta < -0.65 \pi$ the extrapolated gap is very small (of the order of $10^{-3}$ for
$\theta = -0.7 \pi$) which suggests the existence of a phase transition around $-0.67 \pi$
below which the gap is zero. The data shows no evidence for a reopening of the $S=2$ gap in
the interval $\theta \in (-3\pi/4,-0.67\pi]$, at variance with the initial proposal \cite{ChubukovPRB}. 
In order to investigate the closing of the gap by different means we have calculated strong coupling
series expansions of the $S=2$ single particle gap starting from the dimerized limit up to 10th order
in the interdimer coupling $\lambda$ for fixed $\theta$. 
A {\it direct} evaluation of the quintuplet gap shows a closing of the gap around $\theta = -0.67 \pi$ 
which is illustrated in the left panel of Fig.~\ref{fig:serieslambda}.
Using DLog Pad\'e approximants we have determined the critical interdimer coupling $\lambda_c(\theta)$ where
the gap closes, see upper right panel of Fig.~\ref{fig:serieslambda}. Consistent with previous estimates by
Chubukov \cite{ChubukovPRB} we find that the gap closes for the uniformly coupled ($\lambda=1$) chain 
around $\theta \approx -0.67 \pi$.
%%%%%%%%%%%%%%%%%%%%%%%%%%%%%%%%%%%%%%%%%
%% Figure
%%%%%%%%%%%%%%%%%%%%%%%%%%%%%%%%%%%%%%%%%
%%%%%%%%%%%%%%%%%%%%%%%%%%%%%%%%%%%%%%%%%
%% Figure
%%%%%%%%%%%%%%%%%%%%%%%%%%%%%%%%%%%%%%%%%
Having substantiated the closure of the gap in both DMRG and series expansions with indications of an
extended gapless region $-3\pi/4 < \theta < -0.67 \pi$, we now consider the existence of a generalized
BKT phase transition. To determine the critical point we use phenomenological level spectroscopy. 
We calculate the level crossing $\theta_c^L$  of the lowest singlet excitation at $k=\pi$ with the lowest spin-2 
level at $k=0$ for different system sizes $L$ with ED and extrapolate $L \rightarrow \infty$. The results are 
shown in Fig.~\ref{fig:criticalgaptheta}. The extrapolation is performed with $1/L$ and $1/L^2$ corrections, 
fitting only the largest four system sizes. The extrapolated critical point is $\theta_c=(-0.67 \pm 0.02) \pi$, 
in good agreement with estimates obtained by other techniques \cite{Yip,ChubukovPRB}. 
The coefficient of the $1/L$ correction is small, and fitting without this term yields slightly larger 
values $\theta_c > -0.67\pi$.

%%%%%%%%%%%%%%%%%%%%%%%%%%%%%%%%%%%%%%%%%
%% Figure
%%%%%%%%%%%%%%%%%%%%%%%%%%%%%%%%%%%%%%%%%

{\em Dimerization and quadrupolar correlations ---}
Dimerization in the phase diagram of the spin-one chain has been firmly established by exact
results obtained for the $\theta = -\pi/2$ point \cite{DimerizedPhase}. It should be noted that 
although the system is dimerized, it is rather poorly described by a simple product wavefunction
of alternating singlets, as can be seen from the small gap and the large correlation length over 
the whole dimerized phase. The dimerization order parameter \cite{DimerOP} considered here is: 
$(\langle \S_{i-1}\cdot \S_{i}\rangle -  \langle \S_{i}\cdot \S_{i+1}\rangle )$.
The second kind of correlations expected to be important in this region of the phase diagram 
\cite{PapaUnusualPhases,HaradaKawashimaPRB} are the {\em quadrupolar} spin fluctuations 
${\mathcal C}_{\mbox{\tiny quad}}(r)\equiv \langle[(S^z_{0})^2-2/3][(S^z_{r})^2-2/3]\rangle$.

%%%%%%%%%%%%%%%%%%%%%%%%%%%%%%%%%%%%%%%%%
%% Figure
\begin{figure}
  \centerline{\includegraphics[width=0.9\linewidth]{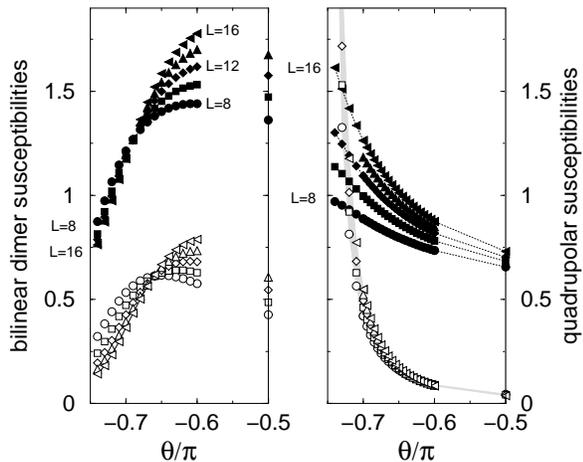}}
  \vspace{-4mm}
  \caption{
    Static and generalized susceptibilities for the bilinear 
    dimerization (left panel) and the ferroquadrupolar correlations (right panel),
    obtained by ED on systems of 8 to 16 sites.
    Solid symbols: Static structure factors.
    Open symbols: Corresponding generalized susceptibilities.
    \label{fig:susceptibilities} 
    \vspace{-5mm}}
\end{figure}
%%%%%%%%%%%%%%%%%%%%%%%%%%%%%%%%%%%%%%%%%

We have calculated the standard static structure factor (SF) and the generalized nonlinear susceptibility (GNS)
\cite{NonlinearSusc} for both kinds of correlations.
We present ED results regarding the dimerization in the left panel of Fig.~\ref{fig:susceptibilities}.
The SF and the GNS both diverge as $L\rightarrow \infty$ for $\theta=-\pi/2$, where long range dimer order
is established \cite{DimerizedPhase}. Similar behavior is found for $\theta \gtrsim -0.67 \pi$.
For $\theta \in (-3\pi/4,-0.67\pi)$ however, we find that the SF and the GNS are both {\em decreasing} with
system sizes, indicating the absence of static dimerization in this gapless region.
In the right panel we display the corresponding quantities for the $k=0$ 
({\em ferroquadrupolar}) mode of the quadrupolar correlations. Here the behavior is different:
while there are only short range correlations deep in the dimerized phase, the ferroquadrupolar 
correlations increase drastically with system sizes for $\theta \in (-3\pi/4,-0.67\pi)$. 
Since quadrupolar long ranged order is forbidden at $T=0$ in this 1D model -- apart from
$\theta=-3\pi/4, +\pi/4$, where the order parameter commutes with the Hamiltonian -- we therefore
expect a power-law behavior. Note that exactly at $\theta=-3\pi/4$ the system is a $SU(3)$ ferromagnet 
and shows ferromagnetism and ferroquadrupolar long range order at the same time \cite{BatistaOrtizGubernatis}.

Finally we want to discuss the relation of the proposed scenario to the numerical results presented in
Ref.~\cite{FathNematic} where an intermediate phase as proposed by Chubukov \cite{ChubukovPRB} could not be found.
We agree with the authors on the absence of a gapped, nematic-like phase. However, we think that the
numerical results in \cite{FathNematic} are compatible with the existence of an intermediate {\em critical},
nondimerized phase, though the authors have not explicitly considered such a phase.

%%%%%%%%%%%%%%%%%%%%%%%%%%%%%%%%%%%%%%%%%
%% Figure
\begin{figure}
  \includegraphics[width=0.8\linewidth]{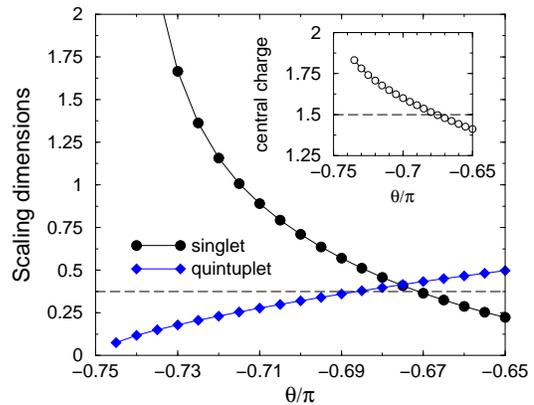} 
  \vspace{-4mm} 
  \caption{
    Conformal field theory parameters calculated in ED on chains
    of 8 up to 18 sites. The scaling dimensions of the singlet ($k=\pi$),
    and the quintuplet ($k=0$) field are shown.
    Inset: effective central charge. The value at the boundary of the
    critical region ($\theta \approx -0.67 \pi$) is $c\approx 3/2$.
    \label{fig:scalingdims}
  \vspace{-7mm}}
\end{figure}
%%%%%%%%%%%%%%%%%%%%%%%%%%%%%%%%%%%%%%%%%

{\em Field theoretical description ---}
In order to discriminate between potential conformal theories describing
the phase transition and the extended gapless region, we calculate relevant 
field theory parameters, i.e.~the scaling dimensions (SD) $x$ of several fields
and the central charge $c$ (Fig.~\ref{fig:scalingdims}). The calculation of these
parameters has been performed with ED and relies on finite size scaling properties 
of groundstate and excited states energies \cite{CFTreferences}. Leading logarithmic
corrections have been taken into account. The SD of the $k=\pi$ singlet field ($x_s$) and 
the $k=0$ quintuplet field ($x_q$) are both close to $3/8$ at the onset of criticality near
$\theta\approx -0.67\pi$. $x_s$ significantly increases, while $x_q$ decreases and seems
to approach $0$ as $\theta \rightarrow -3\pi/4$, compatible with quadrupolar long range order
precisely at $\theta=-3\pi/4$ \cite{BatistaOrtizGubernatis}.
Our results for the effective central charge are unexpected. The central charge $c$ does not seem to be
constant throughout the critical region. The smallest value is found at the onset ($c\approx3/2$),
and then seems to increase monotonously. While this behavior can not directly be ruled out on field
theoretical grounds, it is rather uncommon. It remains to be understood whether $c$ is really 
continuously increasing or whether we are facing a crossover phenomenon. 
The critical theory at $\theta \approx -0.67\pi$ we however believe to be well characterized by a 
level two $SU(2)$ Wess-Zumino-Witten model given both the scaling dimensions ($3/8$) and the central
charge ($3/2$).  Further support for this claim comes from the indirect calculation of $x_s$ within our
series expansions. There the critical exponent $\gamma$ of the $S=2$ gap is related to the scaling 
dimension $x_s$ by $\gamma = 1/(2-x_s)$. In the lower right panel of Fig.~\ref{fig:serieslambda} we show
the critical exponent calculated by DLog Pad\'e approximants to the 10th order series. In the  vicinity 
of $\theta \approx -0.67\pi$ the various approximants reveal only small spreading. Within the precision 
of a 10th order calculation the exponents comply with a scaling dimension of  $x_s = 3/8$ (dashed line).

\begin{figure}
  \centerline{\includegraphics[width=0.8\linewidth]{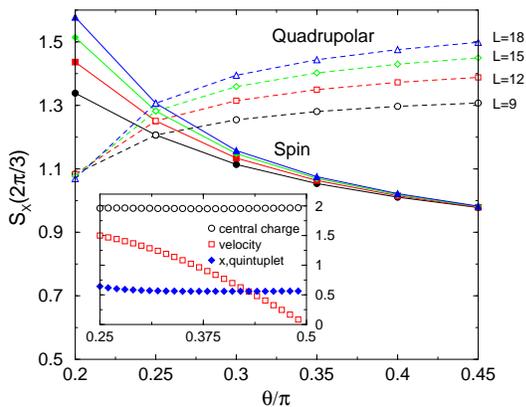}}
  \vspace{-3mm}
  \caption{
    Static structure factors for spin and quadrupolar correlations
    at momentum $k=2\pi/3$ as a function of $\theta$ for different
    system sizes. At the SU(3) point $\theta=+\pi/4$ the two correlation functions
    are symmetry related. By going deeper into the critical region the spin 
    correlations get weaker, while the quadrupolar correlations are enhanced. 
    Inset: the central charge, the excitation velocity and the scaling dimension of the
    quintuplet field in the critical region.
    \label{fig:trimerizedcorrs}
  \vspace{-4mm}}
\end{figure}

{\em The period 3 phase ---}
The existence of an extended critical phase with soft modes at $k=\pm2\pi/3$
has first been reported in \cite{FathPeriodTrippling} and further
analyzed and interpreted in \cite{XiangTrimerized,MuellerKarbach,ItoiKato}.
Our results presented in the first part raise the question whether this
critical region is of spin nematic character as well. Further hints come
from the field theoretical work of Itoi and Kato \cite{ItoiKato}, which shows 
that throughout the critical region the spin 2 mode at $k=\pm2\pi/3$ has correlation-enhancing
logarithmic corrections. This is consistent with the earlier observation,
that the lowest level at $k=\pm2\pi/3$ carries spin 2 \cite{FathPeriodTrippling}.
We have calculated the static spin structure factor ($S=1$) and the spin 
quadrupolar structure factor ($S=2$) both at momentum $2\pi/3$. The results shown in 
Fig.~\ref{fig:trimerizedcorrs} display nicely that the quadrupolar correlations become more 
important than the corresponding spin-spin correlations as one leaves the $SU(3)$ symmetric
point. Although all correlation exponents (dimer, spin, quadrupolar) are equal to $\eta=4/3$,
logarithmic corrections render the quadrupolar correlations dominant in this phase.
Note that this phenomenon is similar to the $S=1/2$ antiferromagnetic Heisenberg
chain, where both spin-spin and dimer-dimer correlations have equal correlation exponents
$\eta=1$, but the logarithmic corrections choose the spin-spin correlations to decay more slowly.
%%%%%%%%%%%%%%%%%%%%%%%%%%%%%%%%%%%%%%%%%
%% Figure
%%%%%%%%%%%%%%%%%%%%%%%%%%%%%%%%%%%%%%%%%
For completeness we show selected field theory parameters for this critical region
in the inset of Fig.~\ref{fig:trimerizedcorrs}. The results are in good agreement with field 
theoretical predictions \cite{ItoiKato} and support the applicability of the used ED techniques 
to determine field theory parameters.

{\em Conclusions and Outlook ---}
To summarize, we have shown that spin nematic correlations play an important role in the phase diagram 
of the $S=1$ bilinear-biquadratic chain. Applying a combination of numerical techniques we reconsidered 
Chubukov's proposal \cite{ChubukovPRB} of a gapped, nondimerized and nematic-like phase between the 
dimerized and the ferromagnetic phases. Our numerical results provide evidence for a modified picture with 
a phase transition at $\theta \approx -0.67 \pi$ where the dimerized phase terminates. However, the phase 
in the window $\theta \in (-3\pi/4,-0.67\pi]$ remains critical, i.e. gapless and nondimerized. The dominant
correlations are spin-quadrupolar correlations at $k=0$. In the second critical region for 
$\theta \in [+\pi/4,+\pi/2)$ we also find dominant quadrupolar correlations at wavevectors $k=\pm2\pi/3$. 
It will be interesting to compare our field theoretical parameters to future analytical calculations.

\acknowledgments
We would like to thank
F.~Alet,
N.~Laflorencie,
P.~Lecheminant,
F.~Mila,
D.~Roggenkamp,
M.~Troyer
and 
S.~Wessel for valuable discussions.
The authors acknowledge support from the Swiss National Science Foundation.


\begin{thebibliography}{}
\bibitem{DimerOP}
  We only discuss the bilinear dimerization, but we have checked that 
  the biquadratic dimerization shows similar behavior regarding the phase boundaries.
\end{thebibliography}
\end{document}